\documentclass{article}
\usepackage{spconf,amsmath,graphicx}
\usepackage{algpseudocode}
\usepackage{algorithm}
\usepackage{multirow}
\usepackage[colorlinks,citecolor=black]{hyperref}
\usepackage{bm}
\usepackage{amssymb}
\usepackage{cite}

\title{Discriminative Speaker Representation via Contrastive Learning with Class-Aware Attention in Angular Space}
\name{Zhe LI $^{1}$, Man-Wai MAK$^{1}$\sthanks{Thanks to Research Grants Council of Hong Kong, Theme-based Research Scheme (Ref.: T45-407/19-N).}{, Helen Mei-Ling MENG$^{2}$} }
\address{$^{1}$Department of Electronic and Information Engineering, The Hong Kong Polytechnic University \\ 
$^{2}$Department of Systems Engineering and Engineering Management, The Chinese University of Hong Kong \\
}

\begin{document}
%
\maketitle
\begin{abstract}

The challenges in applying contrastive learning to speaker verification (SV) are that the softmax-based contrastive loss lacks discriminative power and that the hard negative pairs can easily influence learning. To overcome the first challenge, we propose a contrastive learning SV framework incorporating an additive angular margin into the supervised contrastive loss in which the margin improves the speaker representation's discrimination ability. For the second challenge, we introduce a class-aware attention mechanism through which hard negative samples contribute less significantly to the supervised contrastive loss. We also employed gradient-based multi-objective optimization to balance the classification and contrastive loss. Experimental results on CN-Celeb and Voxceleb1 show that this new learning objective can cause the encoder to find an embedding space that exhibits great speaker discrimination across languages.
\end{abstract}
\section{Introduction}
Speaker verification (SV) requires learning a speaker representation through which utterances are encoded as fixed-sized embedding vectors. Similar speakers will have their embedding vectors close, whereas different speakers will have their embedding vectors far apart. For SV tasks, contrastive representation learning, such as InstDisc \cite{wu2018unsupervised}, SimCLR \cite{chen2020simple}, and MoCo \cite{he2020momentum}, are natural fits to this requirement. These methods aim to discover representations by maximizing the concordance among various augmentations or views of the same instance in a latent space.

\begin{figure}[ht]
\centering
\includegraphics[width=0.5\textwidth]{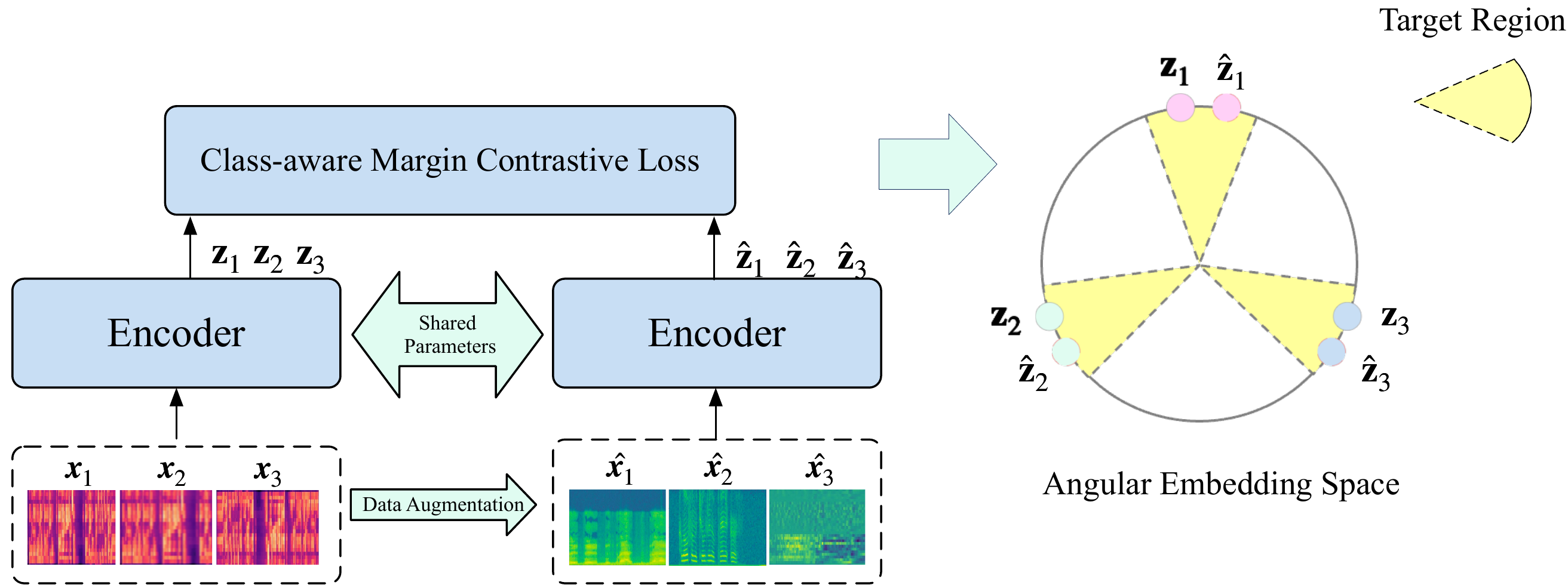}
\caption{Illustration of our basic idea. An embedding space is learned in which the same-speaker pairs stay close to each other while different-speaker pairs remain far apart. For simplicity, we only illustrate three speaker pairs.}
\label{fig:basicidea}
\end{figure}

Contrastive learning has attracted growing attention \cite{tang2021contrastive,zhang2021contrastive,zhang2022contrastive} in the SV community. The technique considers the augmented “views” of an utterance to be spoken by the same person and put them into positive pairs. It treats different utterances and their augmented views to be spoken by different speakers and puts them into negative pairs. The goal is to pull the embeddings in the positive pairs closer together and push those in the negative pairs apart. Many studies \cite{yan2021consert,giorgi2021declutr,gao2021simcse} used the NT-Xent loss, a variant of the cross-entropy loss with a softmax function, to achieve this goal. However, recent studies \cite{wang2018additive,deng2019arcface,li2022real} have demonstrated that the traditional softmax-based loss is effective in optimizing the inter-class difference but not as effective in reducing the intra-class variation. As a result, the learned features are discriminative for closed-set classification but insufficient for open-set speaker recognition. Also, most contrastive approaches focus on determining the positive and negative pairs and improving the learning framework \cite{chen2021self,sang2022self,xia2021self}. The optimization objective is seldom explored. 

A hard sample is defined as one for which the deep learning model struggles to accurately predict the label. However, many sample mining methods used in prior studies \cite{shi2016embedding} do not account for hard negative samples. Therefore, incorporating hard negative samples during training may lead to a higher training loss due to the presence of potential outliers.

To alleviate the above challenges, we propose a framework for learning discriminative speaker representations via \textbf{C}lass-\textbf{A}ware \textbf{A}ttention-based \textbf{Con}trastive learning with \textbf{Margin}, CAAMarginCon. We incorporate an additive angular margin into the supervised contrastive loss, which increases the discrimination capability of the speaker representation by maximizing the decision margin in the angular space. Meanwhile, class-aware attention (CAA) helps identify and ignore very hard negatives. In addition, we employed a multi-objective optimization technique to balance the classification loss and contrastive loss. Fig.~\ref{fig:basicidea} illustrates our idea from a high-level perspective. Experiments conducted on CN-Celeb1\&2 and VoxCeleb1 demonstrate the superiority of the proposed model over the standard ones. In addition, we conducted an ablation study to determine the contributions of different components of the proposed loss function.

\section{Methodology}
\begin{figure*}[ht]
\centering
\includegraphics[width= \textwidth]{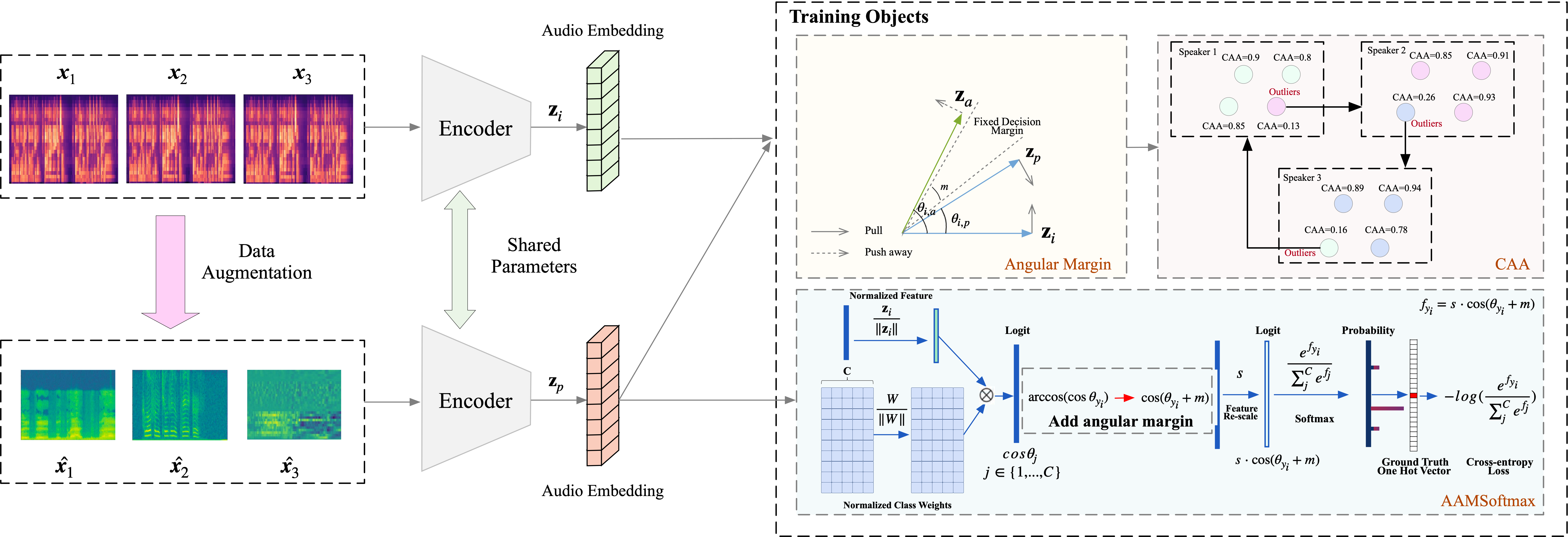}
\caption{The proposed architecture leverages additive angular margin and class-aware attention for supervised contrastive learning. CAA aims to address the impact of hard negative samples.}
\label{fig:CAAMarginCon}
\end{figure*}

Our objective is to train an audio-based feature embedding network using labeled audio. Similar speakers' embedding vectors should be close, whereas those of dissimilar speakers should be far apart. We performed data augmentation on a batch of input samples to produce an augmented batch. Under various augmentations, the embedding vectors derived from the same instance should remain unchanged; on the other hand, the embeddings from different instances should be distinct. As illustrated in Fig.~\ref{fig:CAAMarginCon}, an encoder network receives the spectrograms of both types of instances (the originals and their augmentations) as input and generates normalized embeddings as output. At the network's output, an attention-based class-aware contrastive loss with additive angular margin is computed.
\subsection{Angular Margin-based Contrastive Learning}
We incorporated the original and augmented speaker embeddings into a supervised contrastive loss \cite{khosla2020supervised,li2022speaker}: 
\begin{equation}
\resizebox{\linewidth}{!}{$
    L_{SupCon} = \sum_{i=1}^N \frac{-1}{\vert {\cal P}(i) \vert} \sum_{p \in {\cal P}(i)} \log \frac{\exp(\text{sim}(\bm{z}_i,\bm{z}_p) /\tau)}{\sum_{a \in {\cal A}(i)}\exp(\text{sim}(\bm{z}_i, \bm{z}_a) /\tau)},
$}
\label{Supcon}
\end{equation}
where $\text{sim}(\bm{z}_i, \bm{z}_p)$ is the cosine similarity. In Eq.~\ref{Supcon}, $\bm{z}_i$ is an anchor, $\bm{z}_a$ is a negative sample, ${\cal A}(i)$ is an index set of negative samples with respect to $\bm{z}_i$, $\bm{z}_p$ is a positive sample with respect to $\bm{z}_i$, and ${\cal P}(i)$ contains the indices of positive samples in the augmented batch (original + augmentation). $\tau \in \cal R^+$ is a scalar temperature parameter. Although the training objective attempts to pull the representations of similar speakers closer together and push the representations of different speakers apart, these representations may not be sufficiently discriminative or resilient to noise. 

Let us denote the cosine similarity as
\begin{equation}
\cos{\theta_{i,p}} = \frac{\bm{z}_{i}^{\top} \bm{z}_{p}}{\left\|\bm{z}_{i}\right\| \left\|\bm{z}_{p}\right\|},
\end{equation}
where $\theta_{i,p}$ is the angle between the embedding $\bm{z}_i$ and $\bm{z}_p$. The decision boundary of $\bm{z}_{i} $ for specific $p$ and $a$ is $\theta_{i, p} = \theta_{i, a}$, where $p$ and $a$ are indexes to positive and negative samples, respectively. A tiny perturbation of embedding vectors around the decision boundary may result in an incorrect conclusion if no decision margin is present. To overcome this problem, we propose a new training objective for speaker representation learning by adding an additive angular margin $m$. We name it \textbf{Sup}ervised \textbf{Margin} \textbf{Con}trastive (SupMarginCon) loss, which can be formulated as follows:
\begin{equation}
\resizebox{\linewidth}{!}{$
\begin{aligned}
L&_{SupMarginCon} = \\ & \sum_{i=1}^{N} \frac{-1}{|P(i)|} \sum_{p \in P(i)} \log \frac{\exp \left(\cos \left(\theta_{i, p}+m\right) / \tau\right)}{\sum_{a \in A(i)} \exp \left(\cos \left(\theta_{i, a}\right) / \tau\right)}.
\end{aligned}
$}
\end{equation}
In this loss, the decision boundary of $\bm{z}_{i} $ for specific $p$ and $a$ is $\theta_{i, p}+m=\theta_{i, a}$, which pushes $\bm{z}_{i}$ further towards the area where $\theta_{i, p}$ decreases and $\theta_{i, a}$ increases. Therefore, adding a margin can increase the compactness of same-speaker representations and the disparity between the different-speakers representations. This aid results in improved alignment and uniformity \cite{wang2020understanding} -- two quality measures fundamental to contrastive learning. These measures indicate how close positive-pair embeddings are to one another and how uniformly distributed the embeddings are. In addition, the decision boundary leaves an additional margin $m$ from the boundary $\theta_{i, p}+m=\theta_{i, a}$, which is frequently employed during inference, making the boundary more noise-tolerant and resilient. All of these properties make the SupMarginCon loss more discriminative than the conventional loss, such as the SupCon loss (Eq.~\ref{Supcon}). 

\subsection{Class-Aware Attention}
As indicated in \cite{schroff2015facenet,cui2016fine}, hard samples can easily influence sample mining, resulting in a suboptimal local minimum for the trained model. Usually, we pay more attention to more challenging negatives; however, some of them may be outliers. To determine the relationship between speech embeddings and their classes, we compute the compatibility between an embedding vector and its class vector. The dot product of the two vectors can be used to measure their compatibility \cite{jetley2018learn}.

We apply a classification branch after the speech embeddings to learn the class vectors. The class vectors are trainable and can be updated by gradient descent. Denote $\{\bm{c}_k\}_{k=1}^C$ as the $C$ class vectors. In our case, $C$ is the number of speakers in a mini-batch. For utterances $i$ and $j$, we compute their class-aware attention (CAA) score:
\begin{equation}\label{eq:CAA-score}
\alpha_{i,j} = \frac{\exp(\bm{z}_i^\top\bm{c}_{y_j})}{\sum_{k=1}^C\exp(\bm{z}_i^\top\bm{c}_k)},
\end{equation}
where $y_j$ is the class label of $\bm{z}_j$. Through gradient descent, $\{\bm{c}_k\}_{k=1}^C$ will be able to represent the $C$ classes.

\subsection{Multi-objective Optimization for Weighted Loss}
After finishing the optimization via the contrastive loss, fixing the encoder's parameters is typically required before training a linear classification layer. However, we advocate achieving both contrastive and classification tasks simultaneously. To this end, we introduce AAMSoftmax \cite{deng2019arcface} to our classification task, which is optimized alongside the contrastive loss during training. We employed gradient-based multi-objective optimization \cite{sener2018multi} to perform this muti-task learning. This approach optimizes a proxy objective to minimize a weighted linear combination of losses.

Following AAMSoftmax \cite{deng2019arcface}, we express $\bm{w}_{y_i}^\top\bm{z}_i = \|\bm{w}_{y_i}\|\|\bm{z}_{i}\|\cos\theta_{y_i}$ as the dot product between the class weight of the $y_i$-th training speaker and the embedding $\bm{z}_i$ of the $i$-th utterance. Here, the class weight $\bm{w}_{y_i}$ is the weight vector corresponding to class $y_i$ at the last layer of the classifier, $y_i$ is the speaker label of $\bm{z}_i$, and $\theta_{y_i}$ is the angle between $\bm{w}_{y_i}$ and $\bm{z}_i$. We fix the norm of class weights and embedding vectors to 1 by $l_2$-normalization and re-scale it to $s$. We propose a class-aware attention margin contrastive softmax for supervised embedding learning by combining AAMSoftmax and SupMarginCon:
\begin{equation}
\resizebox{\linewidth}{!}{$
\begin{aligned}
& L_{CAAMarginCon} = \\
&-\frac{\lambda_1}{N} \sum_{i=1}^N \log \frac{\exp(s \cdot \cos(\theta_{y_i}+m))}{\exp(s \cdot \cos(\theta_{y_i}+m))+\sum_{j=1,j \neq y_i}^C \exp(s \cdot \cos\theta_j)} \\ 
& + \lambda_2 \sum_{i=1}^{n} \frac{-1}{|P(i)|} \sum_{p \in P(i)} \log \frac{\exp \left(\cos \left(\theta_{i, p}+m\right) \cdot \alpha_{i,p}/ \tau\right)}{\sum_{a \in A(i)} \exp \left(\cos \left(\theta_{i, a}\right) \cdot \alpha_{i,a} / \tau\right)}.
\label{CAAMarginCon}
\end{aligned}
$}
\end{equation}
where $\alpha_{i,p}$ is the CAA score of pair $(\bm{z}_i, \bm{z}_p)$ and $\alpha_{i,a}$ is the CAA score of pair $(\bm{z}_i, \bm{z}_a)$. By considering the centroid of $\bm{z}_a$ instead of $\bm{z}_a$ itself, we can scale down the contribution of the hard negatives in Eq.~\ref{CAAMarginCon} via the CAA scores in Eq.~\ref{eq:CAA-score}. This is because $\bm{c}_{y_a}$ is {\it less} hard when compared to $\bm{z}_a$. Using the same reasoning, Eq.~\ref{CAAMarginCon} will pay less attention to the {\it easy} positive because $\alpha_{i,p}$ is less than 1.0 even if $\bm{z}_p$ is almost identical to $\bm{z}_i$.
\section{Experiments and Results}
\subsection{Implementation Details}
We adopted 80-dimensional Fbank as the input features. We utilized SpecAugment \cite{park2019specaugment} on the log-mel spectrograms. We augmented the original utterances with noise, music, chatter, and reverberation effects. Our training process was divided into three stages using 200, 400, and 600 frames. The Adam and SGD optimizers were used interchangeably. $m$ is a scalar parameter, and we set the margin $m$ to 0.2. We set the $s$ to 30. The mini-batch size ranges between 512 and 4096. The contrastive learning temperature $\tau$ was set to 0.07.

\subsection{Results and Analysis}
We conducted experiments on CN-Celeb1\&2 and Voxceleb1. For each metric, the best and the second-best are highlighted in bold and underline, respectively. The evaluation results are shown in Table~\ref{mainresult-cnceleb}, from which we can see that CAAMarginCon outperforms the previous approaches. 
\begin{table}[ht]
\caption{The experimental results of the CAAMarginCon and conventional methods on the CN-Celeb evaluation set.}
    \centering
    \resizebox{\linewidth}{!}{
    \begin{tabular}{r|rrr}
\hline
Network & Loss Function & EER(\%) & minDCF \\
\hline
DisSpk-1 \cite{zhu2022discriminative} & Cross Entropy & 11.15 & 0.56 \\
DisSpk-2 \cite{zhu2022discriminative} & Cross Entropy & 11.52 & 0.55 \\
DisSpk-3 \cite{zhu2022discriminative} & Cross Entropy & 11.27 & 0.55 \\
ETDNN\cite{alam22b_odyssey} & Softmax & 10.30 & 0.55 \\
HNN \cite{alam22b_odyssey} & Softmax &  9.18 & \underline{0.50} \\
MSHNN \cite{alam22b_odyssey} & Softmax & 9.05 & \textbf{0.48} \\
ENSEMBLE \cite{alam22b_odyssey} & Softmax   & 8.94 & \textbf{0.48} \\
TDNN-ASP \cite{zeng2022attention} & BCE\&GE2E \cite{wan2018generalized} & 9.90 & 0.56 \\
ResNet34 & RAM-Softmax \cite{li2022real} & 11.05 & N/A \\
ECAPA-TDNN-1 \cite{zeng2022attention} & AAMSoftmax \cite{deng2019arcface} & 10.54 & 0.58 \\
ECAPA-TDNN-2 \cite{zeng2022attention} & AAMSoftmax \cite{deng2019arcface} & 9.51 & 0.56 \\
ECAPA-TDNN-3 \cite{zeng2022attention} & AAMSoftmax \cite{deng2019arcface} & \underline{8.93} & \underline{0.50} \\
ECAPA-TDNN \cite{desplanques2020ecapa} & CAAMarginCon (ours) & \textbf{8.66} & \textbf{0.48}\\
\hline
\end{tabular}
}
\label{mainresult-cnceleb}
\end{table}

Although ENSEMB and MSHNN \cite{alam22b_odyssey} achieve the same minDCF as ours, our EER is lower than theirs. 
Zeng {\it et al.} \cite{zeng2022attention} proposed an excellent attention-based backend by employing scaled-dot self-attention and feedforward self-attention networks that learn the relationships among the enrollment utterances. The different classification results of ECAPA-TDNN in \cite{zeng2022attention} were caused by the fact that this method uses different backends for evaluating similarity. Our proposed model performs better than the discriminative speaker embedding with serialized multilayer multi-head attention in \cite{zhu2022discriminative} because our model is equipped with a discriminative margin and pays less attention to the hard negative samples.

We also explored the performance induced by the CAAMarginCon loss on VoxCeleb1. The results are shown in Table~\ref{result-vox}. We can see that CAAMarginCon enables our network to perform on par with or better than many state-of-the-art systems. 

\begin{table}[ht]
\caption{The experimental results of the proposed CAAMarginCon and conventional methods on the VoxCeleb1 evaluation set. The training set is VoxCeleb1-dev.}
    \centering
    \resizebox{\linewidth}{!}{
    \begin{tabular}{r|rrr}
\hline
Network & Loss Function & EER(\%) & minDCF \\
\hline
MACCIF-TDNN \cite{wang2021maccif} & AAM-softmax \cite{deng2019arcface}& 3.60 & 0.36 \\
ECAPA-TDNN \cite{bai2022end} & Triplet(semi-hard) \cite{schroff2015facenet} & 3.37 & \textbf{0.22}\\
ECAPA-TDNNLite \cite{li2022towards} & AAM-softmax \cite{deng2019arcface}& 3.00 & 0.32 \\
ECAPA-TDNNLite \cite{li2022towards} & AAM-softmax \cite{deng2019arcface} & 2.96 & 0.31 \\
ECAPA-TDNN \cite{desplanques2020ecapa} & AAM-softmax \cite{deng2019arcface}& \underline{2.96} & 0.30\\
ECAPA-TDNN \cite{desplanques2020ecapa} & CAAMargCon (ours) & \textbf{2.85} & \underline{0.29}\\
\hline
\end{tabular}
}
\label{result-vox}
\end{table}
\subsection{Comparing Loss Functions}
To verify the effectiveness of the proposed loss function, we used different loss functions but kept the encoder the same. As seen from Table~\ref{ablation1}, our proposed contrastive loss achieves the best performance, demonstrating the robustness of our proposed model across language. In addition to the cross-entropy loss, CAAMarginCon has a decision margin. It can be seen that the losses with a decision margin are much better than the ordinary softmax (cross-entropy) loss. Compared to AAMSoftmax, CAAMarginCon is superior because of the class-aware attention mechanism. This indicates that a better strategy for mining negative samples can improve the discriminative power of networks trained with contrastive objectives.
\begin{table}[ht]
\caption{The influence of loss functions on CN-Celeb.}
    \centering
\resizebox{\linewidth}{!}{
\begin{tabular}{c|rrr}
\hline
Encoder&Loss Function&EER(\%) &minDCF \\
\hline
\multirow{5}{2.9cm}{ECAPA-TDNN \cite{desplanques2020ecapa}} & Cross Entropy & 16.07 & 0.82 \\ 
& AMSoftmax \cite{wang2018additive} & 13.39 & 0.71  \\
& RAMSoftmax \cite{li2022real} & 13.25 & 0.72 \\
& AAMSoftmax \cite{deng2019arcface} & \underline{8.79} & \underline{0.50}  \\
& CAAMarginCon & \textbf{8.66} & \textbf{0.48} \\ 
\hline
\end{tabular}
}
\label{ablation1}
\end{table}

Table~\ref{ablation2} shows that CAAMarginCon is superior on Voxceleb1. The EER and minDCF of
Softmax (cross-entropy) loss is the highest,
suggesting that the additive angular margin contributes significantly to the performance improvement.

\begin{table}[ht]
\caption{The influence of loss functions on VoxCeleb1.}
    \centering
\resizebox{\linewidth}{!}{
\begin{tabular}{c|rrr}
\hline
Encoder&Loss Function&EER(\%) &minDCF \\
\hline
\multirow{5}{2.9cm}{ECAPA-TDNN \cite{desplanques2020ecapa}} & Cross Entropy & 7.72 & 0.70 \\ 
& AMSoftmax \cite{wang2018additive} & 5.03 & 0.48  \\
& RAMSoftmax \cite{li2022real} & 5.18 & 0.47 \\
& AAMSoftmax \cite{deng2019arcface} & \underline{2.96} & \underline{0.30}  \\
& CAAMarginCon & \textbf{2.85} & \textbf{0.29} \\ 
\hline
\end{tabular}
}
\label{ablation2}
\end{table}

We explored the importance of CAA and margin by removing one component at a time from the CAAMarginCon loss. For simplicity, we set the batch size to 512 and ran for 300 epochs. Table~\ref{ablation3} shows that adding a margin to SupCon can improve performance. Similarly, adding CAA can also improve performance; when the two are combined, the performance is the best.

\begin{table}[ht]
\caption{The influence of CAA and Margin on CAAMarginCon. Results are based on CN-Celeb1\&2 or Voxceleb1-dev for training and CN-Celeb1-test or Voxceleb1-test for evaluations.}
    \centering
\resizebox{\linewidth}{!}{
    \begin{tabular}{c|cccc}
\hline
\multirow{2}{*}{Loss} & \multicolumn{2}{c}{CN-Celeb1-test} & \multicolumn{2}{c}{Voxceleb1-test} \\
 &  EER(\%) & minDCF&EER(\%) & minDCF \\
\hline
CAAMarginCon & 10.31 & 0.61 & 4.06 & 0.46 \\
w/o Margin & 10.22 & 0.64 & 4.15 & 0.46  \\
w/o CAA & 10.39 & 0.62 & 4.11 & 0.46 \\
w/o CAA and Margin & 11.03 & 0.66 & 4.32 & 0.47 \\
\hline
\end{tabular}
}
\label{ablation3}
\end{table}

We also performed experiments on Voxceleb to explore the effect of the CAA and margin. As shown in Table~\ref{ablation3}, both can improve the model's performance. But unlike the performance on CN-Celeb, the effect of the CAA module is not so remarkable. We conjecture that Voxceleb has less noise than the CN-Celeb, i.e., fewer outliers. The CAA module has more contribution to performance under challenging conditions.

\section{Conclusions}
In this work, we propose CAAMarginCon, a supervised contrastive learning framework for learning discriminative speaker representations. CAAMarginCon incorporates an angular margin and a class-aware attention mechanism into the supervised contrastive loss. We optimized the classification and contrastive tasks using a gradient-based multi-objective optimization method. The experimental results of CN-Celeb and VoxCeleb1 demonstrate that both techniques bring substantial improvement.

\bibliographystyle{IEEEbib}
\footnotesize{
\bibliography{strings}
}

\end{document}